\documentclass[aps,prl,showpacs,floatfix,
twocolumn,amsmath,amssymb,superscriptaddress]{revtex4}
\usepackage{graphicx}
\usepackage{color}
\usepackage{ulem}
\usepackage{bm}
\usepackage{hyperref}

\newcommand{\kic}{{\boldsymbol{\kappa}}}
\newcommand{\kicy}{{\kappa_y}}
\newcommand{\Qh}{{h}}
\newcommand{\Qk}{{k}}
\newcommand{\Ql}{{\ell}}
\newcommand{\kpr}{\kappa^{\prime}}
\newcommand{\Hperpa}{${\bf H}\!\parallel\!\boldsymbol{a} $}

\newcommand{\Hperpc}{${\bf H}\!\parallel\!\boldsymbol{c} $}

\newcommand{\Eperpa}{${\bf E}\!\parallel\!\boldsymbol{a} $}

\newcommand{\Eperpc}{${\bf E}\!\parallel\!\boldsymbol{c} $}
\newcommand{\Eperppc}{${\bf E}\!\parallel\!+\boldsymbol{c} $}

\newcommand{\HSFa}{$H_{a}^{\rm SF}$}
\newcommand{\HSFb}{$H_{b}^{\rm SF}$}

\newcommand{\HaGSF}{$H_{a}\! >\!H_{a}^{\rm SF}$}

\newcommand{\Ha}{$H_a$}

\newcommand{\Ea}{$E_a$}

\newcommand{\Ec}{$E_c$}
\newcommand{\MQ}{{\bf M}({\bf Q})}

\newcommand{\SigYpmX}{\sigma_{yx,y\bar{x}}}

\begin{document}
\title{Ferroelectricity from spin supercurrents in LiCuVO$_4$}
\author{M. Mourigal}
\affiliation{Institut Laue-Langevin, BP156, 38042 Grenoble Cedex 9,
                         France}
\affiliation{Laboratory for Quantum Magnetism, \'Ecole Polytechnique
             F\'ed\'erale de Lausanne (EPFL), 1015 Lausanne,
             Switzerland}
\author{M. Enderle}
\affiliation{Institut Laue-Langevin, BP156, 38042 Grenoble Cedex 9,
                         France}
\author{R. K. Kremer}
\affiliation{Max-Planck Institute for Solid State Research,
                         Heisenbergstrasse 1, 70569 Stuttgart, Germany}
\author{J. M. Law}
\affiliation{Max-Planck Institute for Solid State Research,
                         Heisenbergstrasse 1, 70569 Stuttgart, Germany}

\author{B. F\aa k}
\affiliation{Commissariat \`a l'Energie Atomique, INAC, SPSMS, 38054
                         Grenoble, France}
\date{March 22, 2011}
\begin{abstract}
We have studied the magnetic structure of the ferroelectric
frustrated spin-1/2 chain material LiCuVO$_4$ in applied electric
and magnetic fields using polarized neutrons. A symmetry and
mean-field analysis of the data rules out the presence of static
Dzyaloshinskii-Moriya interaction, while exchange striction is shown
to be negligible by our specific-heat measurements. The
experimentally observed magnetoelectric coupling is in excellent
agreement with the predictions of a purely electronic mechanism
based on spin supercurrents.
\end{abstract}
\maketitle

Materials where ferroelectricity (FE) appears simultaneously with
magnetic order are referred to as type-II or magnetic multiferroics.
Considerable attention has been devoted to their unconventional
magnetoelectric (ME) effects \cite{Tokura10}, namely the control of
their electric polarization $\bf P$ with an external magnetic
field~$\bf H$~\cite{Kimura03}, and likewise at $H$=0 the
favoring of a particular magnetic domain by an applied electric
field~$\bf E$
\cite{Radaelli08,Yamasaki07,Seki08,Cabrera09,Finger10}. However, the
fundamental mechanisms responsible for FE in this class of
materials are under debate, in particular whether FE always results
from atomic displacements or whether it can be induced by a purely
electronic mechanism based on spin supercurrents \cite{Katsura05}
via the Aharonov-Casher effect, a relativistic topological quantum
effect for neutral magnetic particles. Here, like in the spin Hall
effect \cite{Murakami03}, the electric field directly controls the
spin current without intermediating atomic displacements.

Phenomenologically, a spin spiral may induce an electric
polarization ${\bf P}\!=\!A\,{\bf e}_{12}\!\times\!({\bf
S}_1\!\times\!{\bf S}_2)$, where ${\bf e}_{12}$ links neighboring
spin sites ${\bf S}_1$ and ${\bf S}_2$ and~$A$ depends on the
fundamental mechanism of the ME coupling~\cite{Mostovoy06}.
Mechanisms that are experimentally established involve symmetry
breaking lattice distortions, which are either induced by symmetric
exchange \cite{Radaelli09} or by asymmetric Dzyaloshinskii-Moriya
interaction (DMI) \cite{Sergienko06,Mochizuki10}. The spin
supercurrent mechanism \cite{Katsura05}, on the other hand, operates
in the absence of magnetostriction and static DMI. Here frustration
of the symmetric exchange interactions leads to non-collinear
magnetic order and a spin supercurrent that couples to the electric
field via the relativistic spin-orbit coupling (SOC). The sign and
size of $A$ is therefore materials specific
\cite{Katsura05,Whangbo07}. However, the spin supercurrent {\it
vanishes} if the spin spiral is stabilized by DMI~\cite{Katsura05}, and the latter is present in many materials, e.g.\ those studied in Refs.~\cite{Yamasaki07,Seki08,Cabrera09,Finger10}.

\begin{figure}
\centering
\includegraphics[width=0.98\columnwidth]{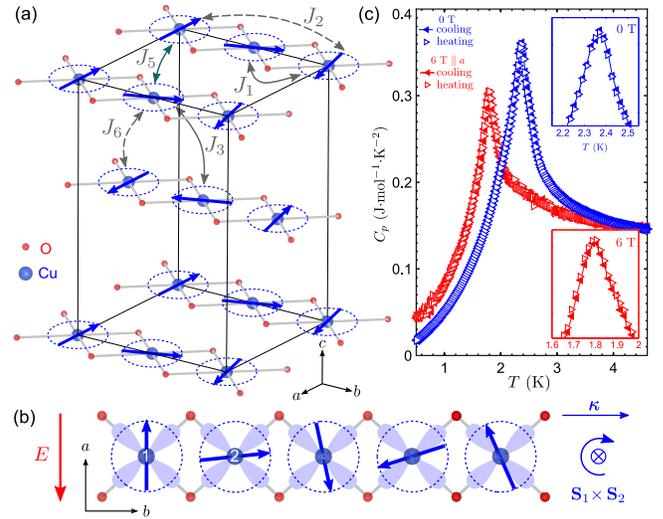}
\caption{(Color online) (a) Unit cell of LiCuVO$_4$ where only
edge-sharing CuO$_4$ chains are shown. Solid (dashed) lines
represent the ferromagnetic (antiferromagnetic) exchange
interactions. (b)~Isolated edge-sharing CuO$_4$ chain with
single-hole $d_{x^2-y^2}$ orbitals and a $-im_b''$ cycloid. (c)~Heat
capacity at the magnetic transition shows absence of hysteresis
between heating and cooling.}
\label{fig1}
\end{figure}

In this Rapid Communication, we provide strong experimental evidence that the
ferroelectricity in the frustrated spin-$1/2$ chain material
LiCuVO$_4$ is purely driven by the spin supercurrent
mechanism~\cite{Katsura05,Whangbo07}, and that magnetostriction is
negligible. By combining mean-field theory and symmetry analysis
with polarized neutron diffraction measurements of the magnetic
structure in crossed electric and magnetic fields, we show that the
static DMI plays no role in driving LiCuVO$_4$ to a spin-spiral
ferroelectric state.

LiCuVO$_4$ crystallizes in the orthorhombic $Imma$ space group
(\#74), where the two Cu$^{2+}$ ions in the primitive unit cell are
located at the positions ${\bf r}_1\!=\!(0,0,0)$ and ${\bf
r}_2\!=\!(0,1/2,0)$ [Fig.~\ref{fig1}(a)]. They form spin--$1/2$
chains along the $b$ axis built up from edge-sharing CuO$_4$ units.
Below $T_N\!=\!2.4$~K, three-dimensional antiferromagnetic (AFM)
order sets in with an incommensurate propagation vector
$\kic\!=\!(0,\kicy,0)$ with $\kicy\!=\!0.532$~\cite{Gibson04},
stabilized by frustration of nearest neighbor ($J_1$) ferromagnetic
(FM) and next-nearest neighbor ($J_2$) AFM exchange interactions
along the chain~\cite{Enderle05,Enderle10}. Simultaneously, an
electric polarization appears along the $a$ axis~\cite{Naito07}. In
a magnetic field along the $a$ axis, a spin-flop transition occurs
at \HSFa$=$2.5~T~\cite{Banks07,Buettgen07}. A concomitant flop of
${\bf P}$ from the $a$ to the $c$ axis is observed for \Hperpa,
while no electric polarization is reported along $b$ irrespective of
the magnetic field direction above \HSFb~\cite{Schrettle08}.

Heat-capacity measurements show that the magnetic order in
LiCuVO$_4$ develops directly from the paramagnetic $Imma$ phase in a
single second-order phase transition without any hysteresis within
the accuracy ($<$5~mK) of the relaxation method of our Quantum
Design calorimeter [Fig.~\ref{fig1}(c)]. This indicates that
symmetry breaking atomic displacements are negligible, which is also
confirmed by our single-crystal neutron diffraction measurements
below $T_N$.

We have determined the zero-field magnetic structure of LiCuVO$_4$
by reanalyzing the unpolarized neutron diffraction data of
Ref.~\cite{Gibson04}. We find that the ordered Cu moments
lie in the $ab$ plane forming a planar spin spiral,
which propagates along the $b$ axis.
This cycloid structure can be
described in terms of Fourier components of the magnetic moments at
the two Cu positions as ${\bf m}_1^\kic = (m_a,\pm i m_b'',0)$ and
${\bf m}_2^\kic = -{\bf m}_1^\kic\exp(-i\pi\kicy)$
with $m_a$ and $m_b''$ being real and positive.
The $\pm$ sign corresponds to two possible
chirality domains.

The population of these two chirality domains can be controlled by
an electric field, as shown by our spherical neutron polarimetry
measurements. In this technique, the directions of the incoming
(${\bf P}_i$) and outgoing (${\bf P}_f$) neutron polarization can be
chosen independently, using a so-called ``cryopad''
device~\cite{CPA}. This device was mounted on the thermal
triple-axis spectrometer IN20 at the Institut Laue-Langevin.
Polarized neutrons of wave vector 2.662~\AA$^{-1}$ were produced and
analyzed by the~$(111)$~reflection of Heusler crystals as
monochromator and analyzer, while second-order (unpolarized)
neutrons were removed by a PG$(002)$ filter in the scattered beam. A
single crystal of LiCuVO$_4$ of size $1.9\times0.7\times5$~mm$^3$
was aligned in the $bc$ plane and cooled down to $T$=1.5 K in a
vertical electric field $E_a \simeq 45$~kV/m pointing in the
$-{\boldsymbol{a}}$ direction. Scans along $(0\Qk0)$ were performed
for magnetic Bragg reflections $(0,\pm\kpr,\pm\Ql)$ with $\Ql\!=\!1,
3, 5,$ and $\kpr\!=\!1\!-\!\kicy\!=\!0.47$.
Introducing a cartesian coordinate system with $\hat{\bf x}$
along~${\bf Q}$, $\hat{\bf y}$ perpendicular to ${\bf Q}$ in the
scattering plane, and $\hat{\bf z}$ perpendicular to the scattering
plane, the neutron cross-section for ${\bf P}_i$ along $\hat{\bf y}$
and ${\bf P}_f$ along $\pm\hat{\bf x}$ is
$\SigYpmX\!\propto\!\frac{1}{2}(|{\bf M}_{\perp{\bf Q}}|^2\!\mp\!i[{\bf
M}^*_{\perp{\bf Q}}\!\times\!{\bf M}_{\perp{\bf Q}}]{\bf \cdot}\bf{\hat{x}})$.
Here, ${\bf M}_{\perp{\bf Q}}\!=\!{\bf Q} \times [\MQ\times {\bf Q}]/|{\bf Q}|^{2}$
is the component of the magnetic structure factor,
$\MQ\!=\!\sum_j {\bf m}_j^\kic\exp(i{\bf Q\cdot r}_j)$,
that is perpendicular to the wave-vector transfer ${\bf Q}\!=\!(\Qh,\Qk,\Ql)$.
We note that the chiral term $i[{\bf
M}^*_{\perp{\bf Q}} \!\times\! {\bf M}_{\perp{\bf Q}}]$ is either
parallel or anti-parallel to {\bf Q}.

\begin{figure}
\centering
\includegraphics[width=0.98\columnwidth]{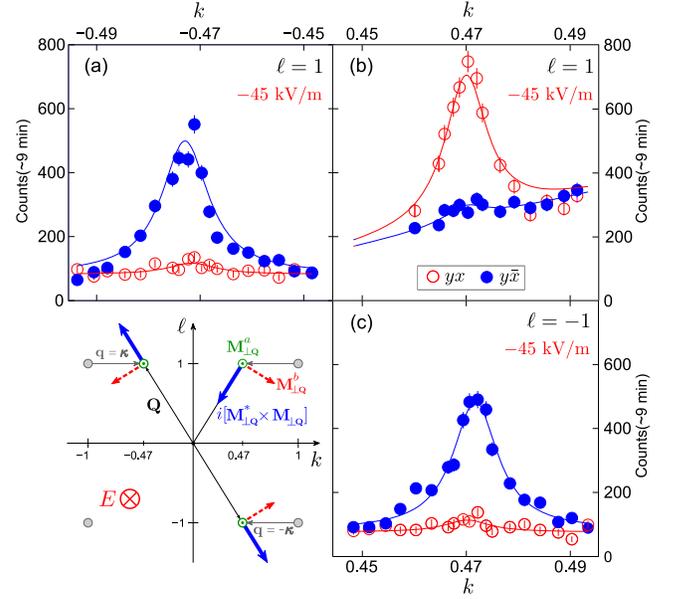}
\caption{(Color online) (a--c) Neutron polarization dependence of
the magnetic reflections $(0,\pm 0.47,\pm1)$ at $T$=1.5 K
with an electric field \Ea~directed along the $-{\boldsymbol{a}}$\ direction.
The lower-left panel is a sketch of the Brillouin zone in the $bc$ plane.}
\label{fig2}
\end{figure}

Figure~\ref{fig2}(a) shows the resulting $\SigYpmX$ cross-sections
for the magnetic Bragg peak $(0,-\kpr,1)$. For the cycloid with $- i
m_b''$ sketched in Fig.~\ref{fig1}(b), the $b$ component of ${\bf
M}({\bf Q})$ is negative (the dashed red arrow shows its projection
$\perp$ {\bf Q}), which implies that the chiral term $i[{\bf
M}^*_{\perp{\bf Q}}\!\times\!{\bf M}_{\perp{\bf Q}}]$ in the
cross-section is parallel to {\bf Q}. Intensity should then be
observed only in the $\sigma_{y\bar{x}}$ channel, which is the case.
At ${\bf Q}\!=\!(0,\kpr,1)$ [Fig.~\ref{fig2}(b)], the direction of
the propagation vector $\kic$ is reversed, which in turn reverses
the $b$ component of ${\bf M}({\bf Q})$, since ${\bf
m}_j^{-\kic}\!=\!({\bf m}_j^\kic)^*$. The sign of the chiral term is
therefore reversed, i.e.\ it is anti-parallel to {\bf Q}, and
intensity is observed in $\sigma_{yx}$, still corresponding to the
same chirality of the cycloid ($- i m_b''$). At ${\bf
Q}\!=\!(0,\kpr,-1)$ [Fig.~\ref{fig2}(c)], the directions of both
{\bf Q} and $\kic$ are reversed, and the situation of
Fig.~\ref{fig2}(a) is recovered. Cooling the sample in zero electric
field leads to equal intensity in the two polarization channels,
which implies that both domains are equally populated. From these
observations we conclude that an electric field \Eperpa\ controls
the chiral domain population and that the ME coupling is negative,
$A\! <\! 0$, since the populated domain has ${\bf
e}_{12}\!\times\!({\bf S}_1\!\times\!{\bf S}_2)$ anti-parallel to
{\bf P} (and {\bf E}).

To obtain more quantitative information on the magnetic structure
for \Eperpa, we modeled all 36 polarized cross sections
$\sigma_{\alpha\beta}$ ($\alpha,\beta\!=\!\pm x,\pm y,\pm z$) of the
five magnetic Bragg peaks measured using a more general ${\bf
m}_1^\kic\!=\!(m_a,m_b'+i m_b'',0)$, i.e.\ allowing $m_b$ to have a
real component in addition to the imaginary one considered until
now. A least-squares fit was performed with $m_a$, $m_b^\prime$,
$m_b^{\prime\prime}$, and $d_-$ as free parameters, where $d_-$ is
the domain population of the $- i m_b''$ domain. The resulting
parameters $m_b^{\prime}\!=\!0.01(5)\,m_a$,
$m_b^{\prime\prime}\!=\!0.95(3)\,m_a$, and $d_-$=0.99(2) establish
that the zero-field magnetic structure is a {\it circular} cycloid
in the $ab$ plane with a single chirality domain fully populated. A
similar fit of two reflections measured with $E$=0 gave an equal
domain population, i.e.\ $d_-$=$d_+$=0.50(0), which strongly
supports the conclusion that full control of the cycloid chirality
is achieved by applying an electric field along $a$.

\begin{figure}
\centering
\includegraphics[width=0.98\columnwidth]{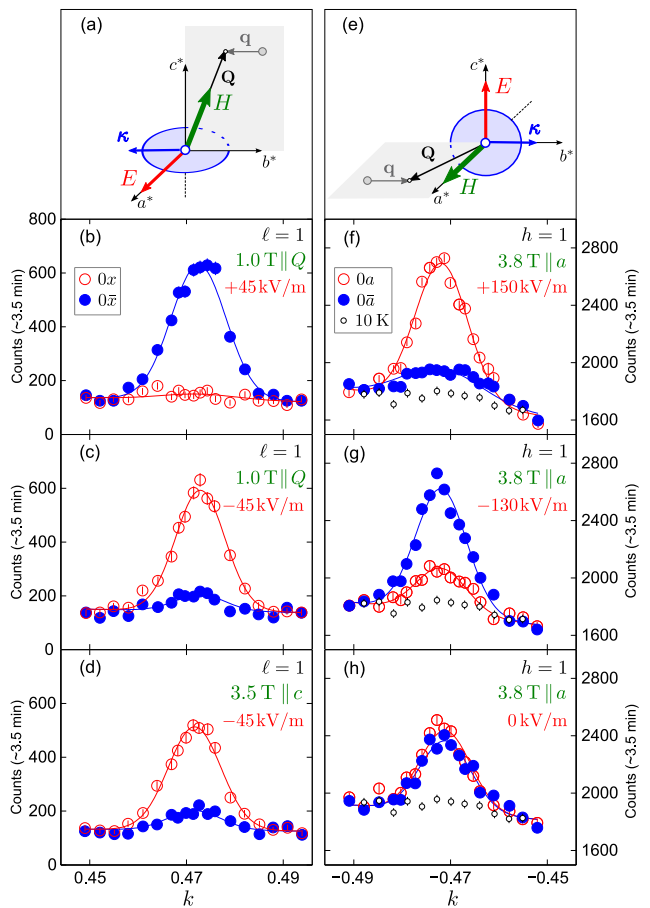}
\caption{(Color online) (a--d) Neutron polarization dependence of
the magnetic reflection $(0,0.47,1)$ at $T$=2.1~K with
\Ea$\simeq\pm45$~kV/m  and $H$=1~T along {\bf Q} as sketched in (a)
or $H$=3.5~T along {\bf c}. (e--g) Similar scans for $(1,-0.47,0)$
at $T$=1.9~K with \Ec~$\neq 0$ and \Ha~=3.8~T along $\boldsymbol{a}$
as sketched in (e). The \Ec$<$0 value had to be reduced with respect
to \Ec$>$0 due to electric discharges. (h) Same scans for $E$=0.
Identical results are obtained with the direction of {\bf H}
reversed.} \label{fig3}
\end{figure}

These results are confirmed when a magnetic field ${\bf H}$ is applied
close to the $c$ axis, in which case no spin-flop transition is
expected nor observed. In order to detect the chiral term we used
the same $bc$-oriented sample as before with a magnetic field in the
horizontal scattering plane and a vertical electric field \Eperpa.
Compared to the previously described set-up, we used an unpolarized
incoming beam from a Si$(111)$ monochromator but kept polarization
analysis and a spin flipper in the scattered beam. In this
configuration, ${\bf P}_f$ is parallel or anti-parallel to {\bf H}.
The neutron polarization for a sample cooled in a positive electric
field is shown in Fig.~\ref{fig3}(b), and, following the same
reasoning as for Fig.\ \ref{fig2}, corresponds to a cycloid with $+
i m_b''$. The sample was then warmed up to $10$~K, well above $T_N$,
and cooled down in a reversed electric field. This leads to a
complete inversion of the chiral domain population
[Fig.~\ref{fig3}(c)]. Complementary measurements with $H$=3.5~T
exactly along the $c$ axis corroborate these findings [see
Fig.~\ref{fig3}(d)]. From this, we conclude that the $ab$ cycloid is
stable upon application of \Hperpc\ up to at least 3.5~T and that an
electric field along $\pm{\boldsymbol{a}}$ fully controls the
chirality of the cycloid. Furthermore, we find that the ME coupling
is negative, $A\! <\! 0$, as in zero magnetic field.

A crucial test of the validity of the spin supercurrent model for
LiCuVO$_4$ is to raise the magnetic field above the spin-flop field
\HSFa, where theory predicts a cycloid in the $bc$ plane, which
according to Ref.\ \cite{Whangbo07} results in a positive ME
coupling, $A\!>\!0$. Measurements were performed on a
$4.5\times3.5\times1$~mm$^3$ single crystal aligned in the $ab$
plane and cooled down with \Hperpa=3.8~T and \Eperppc~as sketched in
Fig.~\ref{fig3}(e). A magnetic peak was found at {\bf
Q}=$(1,\kpr,0)$, which confirms the stability of the propagation
vector $\kic$ for \HaGSF. As shown in Fig.~\ref{fig3}(f--h), ${\bf
P}_f$ changes sign when {\bf E} is reversed (under field-cooled
conditions) and essentially vanishes for zero electric field. For an
electric field ${\bf E}\!\parallel \!+{\bf c}$ ($-{\bf c}$) of 150
(130) kV/m, we observe a circular cycloid given by ${\bf m}_1^\kic =
(0,m,\pm i m)$ with $+i m$ ($-i m$) and a domain population of
$d_+$=0.87 ($d_-$=0.78) [Fig.~\ref{fig3}(f--g)], while for zero
electric field an equal domain population is observed, $d_+$=0.55(2)
[Fig.~\ref{fig3}(h)].
We conclude that an electric
field along the $c$ axis controls the magnetic domain population of the
spin cycloid observed above the spin-flop field, at \Ha$=3.8$~T.
The chirality of the cycloid is such that the ME coupling constant
$A$ is positive for the $bc$-plane structure, $A>0$.

Symmetry analysis, which is directly applicable since there is only
one second order phase transition, leads to four one-dimensional
irreducible representations (IR), none of which can describe the
observed $ab$ cycloid. To proceed, we calculate the mean-field
ground-state energy, $\varepsilon_{\rm gs}(\kic) =
\sum_{\kic}\sum_{ij}[\mathbf{S}_i(-\kic)
\mathbf{J}_{ij}(\kic)\mathbf{S}_j(\kic)]$, where
$\mathbf{J}_{ij}(\kic)$ is a (complex) $6\times6$ matrix
representing the Fourier-transformed exchange interactions. Imposing
the symmetry constraints of the $Imma$ space group on
$\mathbf{J}_{ij}(\kic)$ leads to cancelation of all off-diagonal
terms except terms of type $yz$.
This implies that the DMI term of type $xy$ ($D_z$) is zero for {\it
every} bilinear exchange path, which precludes the stabilization of
an $ab$ cycloid by DMI. A $bc$-cycloid cannot be stabilized either
by DMI, since the $yz$ ($D_x$) term alternates in sign along the
chain. If we impose as constraints our experimental findings that
$h$+$l$ is odd, $m_c$=0, and that the magnetic structure contains
both $m_a$ and $m_b$ components, it follows from minimizing
$\varepsilon_{\rm gs}(\kic)$ that the exchange matrix is
pseudo-tetragonal for all dominant exchange paths defined in Ref.\
\cite{Enderle05}, unless there are pathological coincidences of the
exchanges. The cycloid is then stabilized in the $ab$ plane by a
diagonal exchange anisotropy of easy-plane type,
$|J_z|\!<\!|J_x|\!=\!|J_y|$. Imposing as maximum spin value
$S$=$\frac{1}{2}$ on both Cu sites, we find that the general Fourier
component of site 1, ${\bf m}_1^\kic \!=\! (m_a,\pm [m_b' + i
m_b''],0)$, is reduced to ${\bf m}_1^\kic \!=\! (m,\pm i m,0)$ with
$m$ real. This corresponds to the combination of two 
irreducible representations with equal
amplitude in quadrature and represents two degenerate perfectly
circular cycloids of either rotation sense, in agreement with our
experimental findings. From these mean-field considerations, we
conclude that the zero-field spin-cycloid of LiCuVO$_4$ occurs
without stabilizing DMI, driven exclusively by frustrated
anisotropic diagonal (pseudo-tetragonal) exchange.

A magnetic field applied in the $ab$ plane leads to a spin-flopped
cycloid or helix in a plane perpendicular to the field as soon as
the Zeeman energy overcomes the anisotropic diagonal exchange. Our
finding of a (near) single-domain spin-flopped $bc$-plane cycloid
for \HaGSF\ and \Eperpc\ with the same propagation vector as before
and a chirality that depends on the sign of the electric field
confirms the purely diagonal exchange scheme, and is quite distinct
from the complex behavior of other materials \cite{Mochizuki10}. The
anisotropy of the diagonal exchange energy calculated within
mean-field theory using the spin-flop field of 2.5 T and the main
exchange integrals from Ref.\ \cite{Enderle05} is less than
$\sim$0.5\%. Similar energy considerations for the spin-flop field
put an upper limit on the ratio of off-diagonal ($yz$) and diagonal
exchange energy to less than 4$\cdot 10^{-4}$. The exchange matrix
of LiCuVO$_4$ is hence remarkably isotropic, with diagonal
anisotropic exchange being the leading (small) perturbation. This
pseudo-tetragonal spin symmetry of the exchange matrix has its
origin in the almost tetragonal symmetry of the CuO$_6$ octahedron
and the perfectly flat edge-shared CuO$_4$ chains in the $ab$ plane
of LiCuVO$_4$. The latter lead to a pseudo-tetragonal exchange
matrix even in the presence of a small (orthorhombic) distortion of
the bond angles away from 90$^{\circ}$ \cite{Tornow99}.

By combining our experimental results with mean-field calculations,
we have shown that the formation of a ferroelectric spin-cycloid in
LiCuVO$_4$ below $T_N$ is driven by purely diagonal frustrated
anisotropic exchange without stabilizing static DMI or
magnetostriction. This rules out most of the currently discussed
mechanisms for FE except the spin supercurrent scenario
\cite{Katsura05}.
This purely electronic scenario has been studied for both linear
Cu--O--Cu clusters of $e_g$ electrons~\cite{Jia07} and for
edge-sharing CuO$_4$ geometry \cite{Whangbo07,Tornow99,Onoda07}. In
the latter case, SOC has been shown to generate a small symmetric
exchange anisotropy \cite{Tornow99} as well as Coulombic
ring-exchange processes \cite{Tornow99,Onoda07}, which lead to
chiral correlations and finally ME coupling \cite{Onoda07}. Tetramer
diagonalization and first-principles calculations of the SOC on O-
and Cu-sites of LiCuVO$_4$ show that the spin supercurrent creates
FE polarization in the absence of atomic displacements, only by
asymmetric orbital occupancy \cite{Whangbo07}. These calculations
predict the sign change of the ME constant $A$ at the spin-flop
transition for \Hperpa\ that we observe experimentally.

We conclude that the purely electronic scenario of spin
supercurrents is realized in LiCuVO$_4$: a small symmetric exchange
anisotropy is present while magnetostriction and static DMI are
negligible, and the theoretical predictions of both the ratio of the
electric polarizations $P_a/P_c$ and the sign of the ME coupling in
all ferroelectric phases are found in excellent agreement with
experiments.

We acknowledge fruitful discussions with J.~Schweizer, G.~Nenert,
T.~Ziman, and M.~E.~Zhitomirsky and experimental assistance from
P.~Steffens, P.~Chevalier, E.~Bourgeat-Lami (neutrons), and
G.~Siegle (specific heat).

\end{document}